\documentclass[runningheads]{llncs}
\usepackage{graphicx}
\usepackage{subfigure}
\usepackage{booktabs}
\usepackage[marginal]{footmisc}

\begin{document}
\title{Self-Adaptive Transfer Learning for Multicenter Glaucoma Classification in Fundus Retina Images}
%
%
\author{Yiming Bao\inst{1}, Jun Wang\inst{1}, Tong Li\inst{1}, Linyan Wang\inst{2}, Jianwei Xu\inst{1}, Juan Ye\inst{2} and Dahong Qian\inst{1}}
\authorrunning{Y. Bao et al.}
\titlerunning{Self-Adaptive Transfer Learning}
%
\institute{Institute of Medical Robotics, Shanghai Jiao Tong University, Shanghai, China 
\email{dahong.qian@sjtu.edu.cn}\and
the Department of Ophthalmology, the Second Affiliated Hospital of Zhejiang University, College of Medicine, Hangzhou, China}
\maketitle              
\begin{abstract}
The early screening of glaucoma is important for patients to receive treatment in time and maintain eyesight. Deep learning (DL) based models have been successfully used for computer-aided diagnosis (CAD) of glaucoma. However, a DL model pre-trained on certain dataset from one hospital may have poor performance on other hospital data, therefore its applications in the real scene are limited. In this paper, we propose a self-adaptive transfer learning (SATL) strategy to fill the domain gap between multi-center datasets. Specifically, the encoder of a DL model that is pre-trained on the source domain is used to initialize the encoder of a reconstruction model. Then, this reconstruction model is trained using only unlabeled image data from the target domain, which makes the encoder in the model adapt itself to extract useful features both for target domain images encoding and glaucoma classification, simultaneously. Experimental results on a private and two public glaucoma diagnosis datasets demonstrate that the proposed SATL strategy is effective. Also, it meets the real scene application and the privacy protection policy due to its independence from the source domain data.

\end{abstract}
\footnote{\noindent Y. Bao and J. Wang are co-first authors. J. Ye and D. Qian are co-corresponding authors}
\section{Introduction}
\label{sec:introduction}
Glaucoma is one of the most primary leading causes of blindness\cite{mary2016retinal}. The loss of sight due to glaucoma is irreversible while some other eye diseases such as myopia and presbyopia are not. Thus, early diagnosis of glaucoma for effective treatment and vision conservation matters a lot for patients. 

However, the symptoms of glaucoma in the early stage are difficult to perceive. One of the standard methods widely used by eye specialists nowadays is the optic nerve head (ONH) assessment\cite{mary2016retinal} in fundus retina images. Whereas, mastering the tricks of performing ONH assessment remains challenging. Therefore, some automatically calculated parameters were presented and popularized as quantitative clinical measurements, such as cup to disc ratio (CRD) which means the ratio of vertical cup diameter to vertical disc diameter in the fundus retina image. Generally, a larger CRD represents a higher possibility of glaucoma and vice verse. However, manually labeling the mask of the cup or disc region is labor-consuming, which makes image-level category labels necessary and reasonable for automatically screening glaucoma.

In the past several years, Deep Learning (DL) based methods have received unprecedented attention and achieved state-of-the-art performance in many fields, including medical image analysis\cite{ravi2017deep}. Glaucoma can be screened from fundus retina images by DL models which are well trained on sufficient data and precise image-level labels\cite{fu2018disc-aware}. However, DL models trained on one single site cannot be directly generalized and applied to other sites. The distributions of training and testing data are partially different so the pre-trained model may fail to fulfill the diagnosis task.

Commonly, the difference between datasets can be seen as a domain gap. For Example, the discrepancy between images from different dataset can be reflected in many image statistical traits, such as color style, contrast, resolution, and so on. Also, the joint distributions of images and labels may be quite different between the source and the target domain, i.e., $P(x^s,y^s ) \neq P(x^t,y^t )$. This is mainly because the margin distributions are different, i.e., $P(x^s)\neq P(x^t)$ even if the conditional distributions, i.e., $P(y^s\vert x^s)$ and $P(y^t\vert x^t)$ are similar. Many methods have been proposed to solve this problem. Fine tuning\cite{tajbakhsh2016convolutional} is most widely used in real practical applications. However, fine-tuning is unable to apply when the dataset from a new target domain is completely unlabeled. 

To solve the domain adaptation problem, a novel \emph{self-adaptive transfer learning} (SATL) framework is proposed in this paper for glaucoma diagnosis. Specifically, we train a convolutional neural network in the source domain with sufficient labeled data. Then, the feature extraction layers of this trained model is shared as the encoder of a reconstruction network. The reconstruction network is trained in the target domain using only unlabeled data. The encoder is adapted to fit the distribution of target data while maintains the ability for glaucoma diagnosis. The contributions of this paper can be concluded as follows:

(1) To the best of our knowledge, our work is the first to investigate the study of transfer adaptation learning for the classification of glaucoma with multicenter fundus retina images. 

(2) Our framework only uses unlabeled date in the target domain and is independent from source domain data, so it has great potential for real scene applications and can meet privacy protection policy for medical data.

(3)  Experimental results shows that our framework can preserve most of the classification ability of the off-shelf model and meanwhile improve its classification performance in target domain data. Even totally independent from source domain data, it outperforms other state-of-the-art domain adaptation methods such as CycleGAN, which heavily relies on source domain data in adaptation stage. 

\section{Related Works}

Transfer adaptation learning(TAL)\cite{zhang2019transfer,wang2018deep} is the most relevant area with the proposed method. It is a combination of transfer learning (TL) and domain adaptation (DA) and can be categorized into three classes, which will be introduced respectively.

\textbf
{Instance Re-weighting Adaptation Learning (IRAL)}
Methods in this area assign weights to the source domain instances based on their similarity to the target domain instances \cite{zhu2020boundary-weighted,qi2019label-efficient}. Via re-sampling or importance weighting, the performance of the trained source classifier in the target domain can be enhanced. However, the estimation of the assigned weights is under a prior-decided parametric distribution assumption\cite{zhang2019transfer}, which may differ from the true parametric distribution. 

\textbf
{Feature Adaptation Learning (FAL)}
For adapting datasets from multiple domains, methods in this category are widely proposed to find a feature representation space where the projected features from target and source domain follow similar distributions \cite{wang2019patch-based,shen2020domain-invariant}. In the past few years, the most famous FAL methods are GAN-based domain adaptation models. However, finding a general feature space for most domains remains challenging. Also, training a GAN-based domain adaptation model needs both source and target domain data, which is more and more impractical in the real scene due to the privacy protection policy for medical data. 

\textbf
{Self-Supervised Transfer Learning (SSTL)}
Algorithms in this category focus on training a supervised classifier on the source domain and then transfer its knowledge to the target domain via self-supervised learning \cite{cheng2015multimodal,cheplygina2018transfer,ghifary2016deep,sun2020gan}
. For example,  Cheplygina \emph{et al}\cite{cheplygina2018transfer} investigated a Gaussian texture features-based classification model of chronic obstructive pulmonary disease(COPD) in multicenter datasets. These methods integrate the data information from different domains by extracting some manually designed features from images, which limits the generalization ability of model. Ghifary \emph{et al} \cite{ghifary2016deep} is the most relative literature with our framework. Our method differs from \cite{ghifary2016deep} mainly in the network structure. Moreover, we explore application in glaucoma diagnosis in several datasets.

\begin{figure*}[!ht]
\begin{center}
    \centerline{\includegraphics[width=\linewidth]{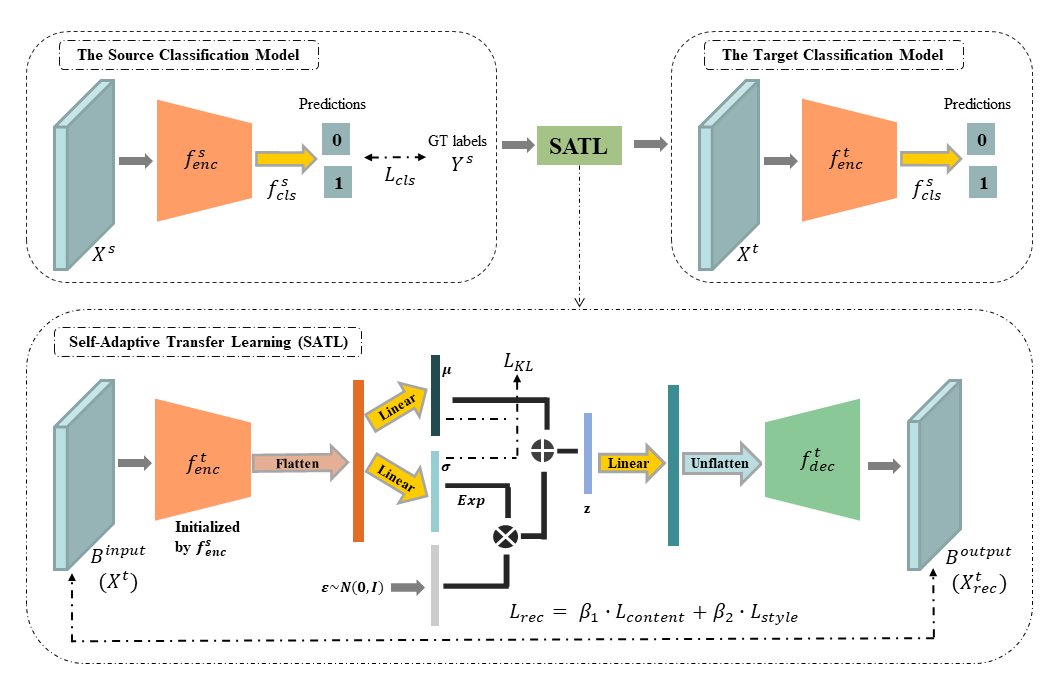}}
\caption{Illustration of the self-adaptive transfer learning (SATL) strategy, which is independent of the source domain data and more suitable for the real scene applications.}
\label{fig1}
\end{center}
\end{figure*}

\section{Method}

The framework of the proposed method is illustrated in Fig. \ref{fig1}. The proposed SATL framework can transfer a pre-trained source classification model to a target domain without using neither source images nor labels.

Let $f^{s}: \mathcal{X}^{s} \to \mathcal{Y}^{s}$ be the source pre-trained classification model and $f^{t}: \mathcal{X}^{t} \to \mathcal{X}^{t}_{rec}$ be the target reconstruction model. The feature encoder is denoted as $f_{enc}:\mathcal{X} \to \mathcal{F}$ and the lightweight classification function $f_{cls}: \mathcal{F} \to \mathcal{Y}$. We denote one more function: an decoder  $f_{dec}:\mathcal{F} \to \mathcal{X}$ in $f^{t}$. Then, given an input sample $x$, $f^{s}$ and $f^{t}$ can be formulated as :

\begin{equation}
    f^{s}(x) = f^{s}_{cls}(f^{s}_{enc}(x)); f^{t}(x) = f^{t}_{dec}(f^{t}_{enc}(x))
\end{equation}

Once $f^{t}(x)$ is trained, we can build the self-adapted classification model
$f^{t}_{SA}(x)$ for target domain image classification by $f^{t}_{SA}(x) = f^{s}_{cls}(f^{t}_{enc}(x))$
 
As shown in Fig. \ref{fig1}, the reconstruction model $f^{t}_{dec}$ is implemented as a variational auto-encoder (VAE), which can compress the image information and sample a latent vector $z$. The encoder of it $f^{t}_{enc}$ is initialized by the pre-trained source encoder $f^{s}_{enc}$.

The loss function used to optimize the proposed self-adaptive reconstruction model can be represented as:

\begin{equation}
    L(f^{t}_{enc},f^{t}_{dec},x^{t})  = \alpha \cdot L_{KL} + \beta \cdot L_{rec},
\end{equation}

\begin{equation}
    L_{KL} =  - KL(f_{enc}^{t}(z|x^{t})|f_{dec}^{t}(z|x^{t})), 
\end{equation}

where the first term in the loss function $L_{KL}$ is the KL divergency of the latent vector distribution and the true data distribution. The second term $L_{rec}$ is the reconstruction loss between the output image and the input image. Instead of using a single MSE loss, we perform a new designed combination of two loss functions following \cite{li2017demystifying}. We argue that the self-adaptive reconstruction model should be guided to reconstruct high-level style information in the target domain images rather than just the pixel-wise texture. Thus, the reconstruction loss function designed in this paper is as:

\begin{equation}
    L_{rec}  = \beta_1 \cdot \sum_{i,j,k}(B^{output}_{ijk} - B^{input}_{ijk})^2 + \beta_2 \cdot \sum_{m,n}(G^{output}_{mn} - G^{input}_{mn})^2,
\end{equation}

where $B^{output}$ and $B^{input}$ denote the output and input of the reconstruction model, respectively. $i,j,k$ and $m,n$ represent the position indexes.  $G^{output}$ and $G^{input}$ are the Gram matrices of $B^{output}$ and $B^{input}$. The gram matrix can be calculated as:

\begin{equation}
    G = \frac{1}{n_i \times n_j \times n_k} \mathbf{v} \mathbf{v^{T}},
\end{equation}

where $\mathbf{v}$ is the flattened column vector of $B^{output}$ or $B^{input}$.

\section{Experiments and Results}

\subsection{Datasets}

\begin{table}
\centering
\caption{The statistical difference between three datasets}
\begin{tabular}{c|c|c|c|c}
\toprule
Dataset& 
Domain&
Samples &
Pos vs. Neg &
Avg of image size\\
\hline
LAG (public)&
Source / Target&
4854&
3143:1689&
$300 \times 300$\\
\hline
pri-RFG (private)&
Source / Target&
1881&
$1013:868$&
$989 \times 989$\\
\hline
REFUGE (public)&
Target only&
400&
$40:360$&
$1062 \times 1062$\\
\bottomrule
\end{tabular}
\label{tab1}
\end{table}

We used two public datasets and one private dataset to validate the proposed SATL framework on glaucoma diagnosis task. The first public dataset is large-scale attention-based glaucoma(LAG) dataset\cite{li2019attention} established by Li \emph{et al}. The second is from the REFUGE challenge\cite{orlando2020refuge}. Moreover, we also collected 1881 retina fundus images from one collaborated hospital and built a private dataset (pri-RFG) via labeling all the images by experienced ophthalmologists. The details of the above-mentioned three datasets (LAG, REFUGE, pri-RFG) are summarized and tabulated in Table \ref{tab1}. We can observe that the scales, the average size of images and the ratio of samples in different datasets are quite various, making transfer learning between them challenging. Due to the small number of samples in dataset REFUGE, we just used it as target domain dataset, while LAG and pri-RFG are used for cross-domain evaluation. In other words, we implemented a total of four groups of experiments. Based on the direction from source domain to target domain, they can be represented as LAG $\to$ pri-RFG, pri-RFG $\to$ LAG, LAG $\to$ REFUGE and pri-RFG $\to$ REFUGE. When used as a source domain dataset, we separated training and validation set. When used as a target domain dataset, all the images were fed into the reconstruction model to train and adapt the encoder layers.

\subsection{Implement Details and Evaluation Metrics}\label{3b}

Both the source classification model and the target reconstruction model were implemented using Pytorch (version 1.3.0) and trained on an NVIDIA RTX 2080Ti GPU. We implemented the source classification model as a VGG\cite{simonyan2014very} and optimized it with cross entropy (CE) loss\cite{ng2001lag}. During the training stage of the source classification model, we set the learning rate as $10^{-6}$, weight decay as $5 \times 10^{-4}$. All the samples in the source domain were split into training set and validation set using a ratio of $7:3$ empirically, following stratified sampling method to ensure that the Pos vs. Neg ratios in each set are similar. At each iteration, a mini-batch of 16 samples were fed into the model. The number of training epochs was set as 50. To avoid the over-fitting issue, the model which achieved the maximum accuracy in the validation set was saved.

During the training stage of the self-adaptive reconstruction model on the target dataset, the learning rate of the encoder was set as $10^{-7}$ and that of the rest layers was set as $10^{-3}$. To avoiding over-fitting on the reconstruction task and losing the ability to extract features that are useful for classification task, the target reconstruction model was trained for only 20 epochs. We empirically set the weights $\alpha$, $\beta_1$ and $\beta_2$  in the reconstruction loss function as 0.3, 0.2, 0.5, and the channel number of the latent vector in the model as 32.

Once the target reconstruction model was trained, the self-adapted encoder of it was used as the feature extractor of a target classification model. The last lightweight FC layer of the source classification model played a role as classifier. This new combined target classification model was evaluated on target domain dataset by metrics in terms of Accuracy, Recall, Precision, F1 score and Area Under the ROC Curve (AUC).

\subsection{Results and discussion}

As described in Section \ref{3b}, based on the three available datasets, there are four executable domain adaptation directions denoted as LAG $\to$ pri-RFG, pri-RFG $\to$ LAG, LAG $\to$ REFUGE, and pri-RFG $\to$ REFUGE. For validating the effectiveness of the proposed SATL strategy, on each experiment direction we compared the performance of proposed method (\textbf{w/ SATL}) with the source classification model (\textbf{w/o SATL}) and a state-of-the-art CycleGAN-based domain adaptation method\cite{zhu2017unpaired} (\textbf{w/ CGAN}). The CycleGAN-based method trains a generator to transfer the target images to the source domain by adversarial learning. The most noteworthy difference between CycleGAN and the proposed SATL strategy is that: our method is completely independent of the source domain data while CycleGAN is not. More specifically, training CycleGAN to perform domain adaptation needs both source and target domain images. On the contrary, the proposed SATL strategy relies on only the target domain unlabeled images. 

\begin{table}
\centering
\caption{The classification performance of four groups of experiments}
\begin{tabular}{c|c|c|c|c|c|c}
\toprule
Direction& \multicolumn{3}{c|}{LAG $\to$ pri-RFG}
& \multicolumn{3}{c}{pri-RFG $\to$ LAG}\\
\hline
Strategy& w/o SATL & w/ CGAN &w/ SATL
& w/o SATL & w/ CGAN & w/ SATL\\
\hline

Accuracy &0.799 &0.672 &\textbf{0.856} 
&0.352 &\textbf{0.628} &0.579 \\

Recall &0.659 &0.422 &\textbf{0.726}
&\textbf{1.000} &0.707 &0.779\\

Precision &0.807 &\textbf{0.923} &0.855 
&0.352 &\textbf{0.481} &0.445\\

F1 Score &0.726 &0.580 &\textbf{0.785} 
&0.521 &\textbf{0.573} &0.566\\
\hline
Direction& \multicolumn{3}{c|}{LAG $\to$ REFUGE}
& \multicolumn{3}{c}{pri-RFG $\to$ REFUGE}\\
\hline
Strategy& w/o SATL& w/ CGAN  & w/ SATL
& w/o SATL & w/ CGAN & w/ SATL\\
\hline
Accuracy &0.933 &0.913  &\textbf{0.945} 
&0.240 &0.540 &\textbf{0.580}\\

Recall &0.425 &\textbf{0.600} &0.500
&\textbf{0.975} &0.825 &0.850\\

Precision &0.810 &0.558  &\textbf{0.909} 
&0.114 &0.157 &\textbf{0.173}\\

F1 Score &0.557 &0.579  &\textbf{0.645} 
&0.204 &0.264 &\textbf{0.288}\\
\bottomrule
\end{tabular}
\label{tab2}
\end{table}

\begin{figure}[!ht] 
\begin{center}
    \centerline{\includegraphics[width=\linewidth]{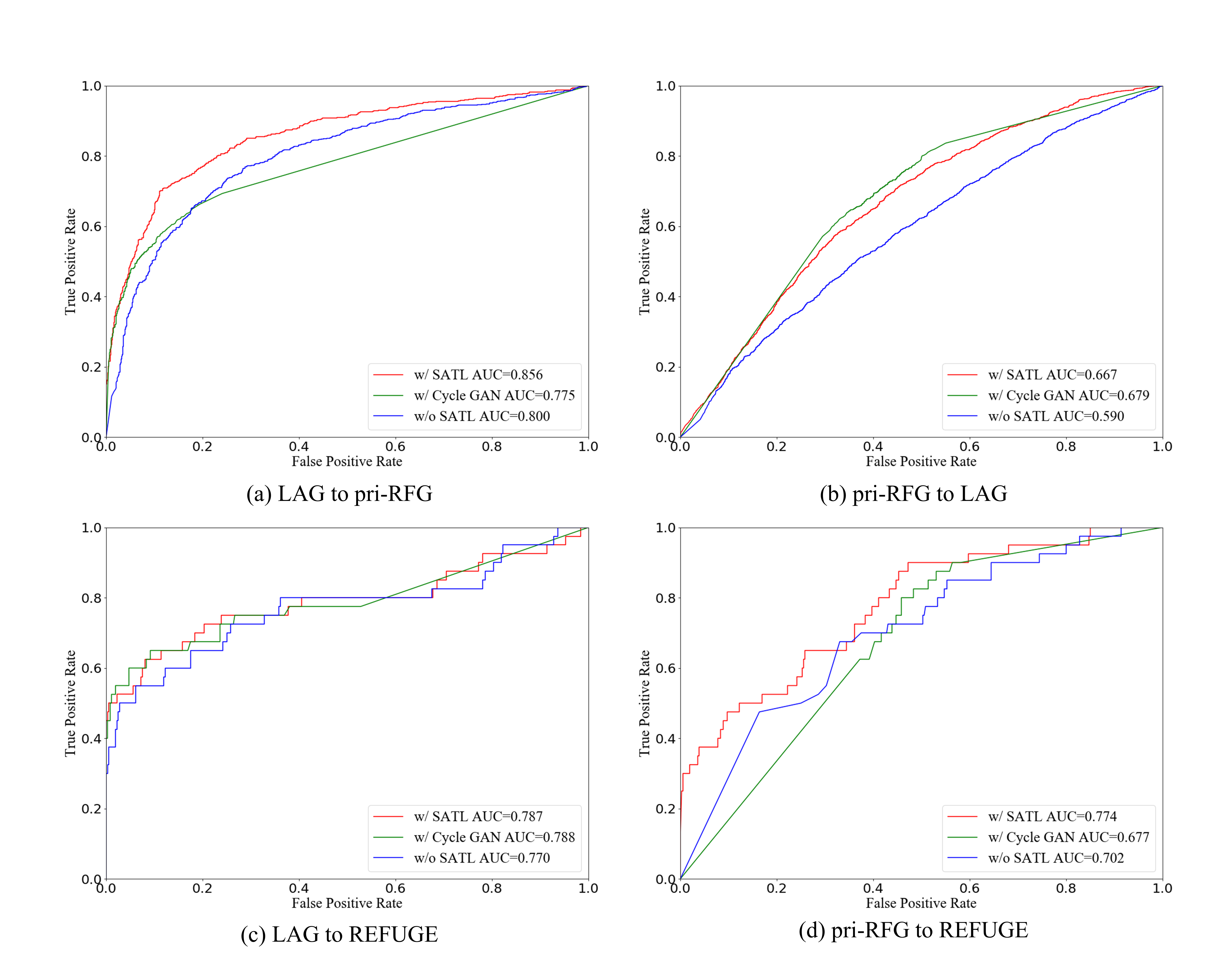}}
\caption{ROC curves of the models evaluated in all four domain adaptation directions.}
\label{fig2}
\end{center}
\end{figure}

The experimental results of three strategies are tabulated in Table \ref{tab2}. Moreover, the ROC curves are also plotted and illustrated in Fig. \ref{fig2}. By observing the demonstrated results, two main conclusions can be drawn: 

(1) Compared to the source model without SATL, which can be seen as a baseline, the model with SATL outperforms in all four domain adaptation directions in terms of Accuracy and F1 Score. Despite there exist a mass of differences between three used datasets, SATL shows to be effective for self-supervised domain adaptation regardless of the source and target domain data distribution. This phenomenon shows that the proposed SATL is valuable and reliable for the production of pseudo labels in data from a grand-new hospital.

(2) When testing the source model in the target domain images transferred by CycleGAN, the performance is comparable with the proposed SATL strategy in domain adaptation directions of pri-RFG $\to$ LAG and LAG $\to$ REFUGE. While in directions of LAG $\to$ pri-RFG and pri-RFG $\to$ REFUGE, the proposed SATL strategy surpasses the CycleGAN by a large margin. This phenomenon demonstrates that SATL is more robust and have more stable generalization ability in different domain adaptation scenes. Note that CycleGAN uses the source domain images in the domain adaptation stage while the proposed SATL does not. Thus, our method which is completely independent of the source domain is more feasible in real scene applications. It can ensure the isolation of multi-center datasets and meet the privacy protection policy.

\textbf{Discussion} Despite the proposed method improves the performance of the classification model in the target domain via self-supervised training, there still remains some research worth exploring for enhancing the performance. For example, in this paper, we directly trained and validated the source classification model on the source domain. However, it may be a better option to initialize the source classification model by a model pre-trained on large scale nature image datasets such as ImageNet. Besides, the backbone used in this paper is VGG for the convenience of building the reconstruction VAE model. In the future, it can also be replaced by other state-of-the-art backbone such as Inception\cite{szegedy2016inception-v4} or SENet\cite{hu2019squeeze-and-excitation}. Last but not least, the features adapted by SATL framework in the target domain need to be explore and compare with that 
before SATL. Further improvement in glaucoma diagnosis may be achieved by learning features which can better represent ONH traits.

\section{Conclusion} In this paper, we present a self-adaptive transfer learning (SATL) strategy to fill the domain gap between multicenter datasets and perform the evaluation in glaucoma classification based on three fundus retina image datasets. Specifically, a reconstruction model is trained using only target domain unlabeled images. The encoder of this reconstruction model is initialized from a pre-trained source classification model and self-adapted in the target domain. Experimental results demonstrate that the proposed SATL strategy enhances the classification performance in the target domain and outperforms another state-of-the-art domain adaptation method which even utilizes source domain images for training, as well. In the near future, more efforts will be devoted to exploring how to furthermore lifting the performance of the self-supervised domain adaptation method via designing new reconstruction losses. Moreover, we will extend this strategy to other medical image analysis problems.

\textbf{Acknowledgement.} This work was supported in part by Department of Science and Technology of Zhejiang Province - Key Research and Development Program under Grant 2017C03029 and the Biomedical Engineering Interdisciplinary Research Fund of Shanghai Jiao Tong University under Grant YG2020YQ17.
%
%
%
%
{\small
\nocite{*}
\bibliographystyle{splncs04}
\bibliography{egbib}
}





\end{document}